\documentstyle[epsfig]{aipproc}

\begin{document}
\title{Laser spectroscopy of simple atoms
and precision tests of bound state QED}

\author{Savely G. Karshenboim\thanks{E-mail: sek@mpq.mpg.de}}
\address{D.I. Mendeleev Institute for Metrology, 198005 St. Petersburg, Russia\\
Max-Planck-Institut f\"ur Quantenoptik, 85748 Garching, Germany\thanks{Summer address}
}

\maketitle

\begin{abstract}
We present a brief overview of precision tests of bound state QED and 
mainly pay our attention to laser spectroscopy as an appropriate tool 
for these tests. We particularly consider different precision tests of  
bound state QED theory based on the laser spectroscopy 
of optical transitions in hydrogen, muonium and positronium 
and related experiments. 
\end{abstract}


\section{Introduction}

Precision laser spectroscopy of simple atoms (hydrogen, deuterium, muonium, 
positronium {\em etc.}) provides an opportunity to precisely test Quantum 
Electrodynamics (QED) for bound states and to determine some fundamental 
constants with a high accuracy. The talk is devoted to a comparison of theory and 
experiment for bound state QED.
Experimental progress during the last ten years has been mainly due to laser spectroscopy 
and, thus, the tests of bound state QED are an important problem associated with modern laser physics.

The QED of free particles (electrons and muons) is a well-established theory designed to 
perform various calculations of particle properties (like e. g. anomalous magnetic moment) 
and of scattering cross sections. In contrast, the theory of bound states is not so well 
developed and it needs further precision tests. 
The QED theory of bound states contains three small parameters, which play a key role: 
the QED constant $\alpha$, the strength of the Coulomb interaction $Z\alpha$  and the mass ratio 
$m/M$  of an orbiting particle (mainly---an electron) and the nucleus. It is not possible to do any 
exact calculation and one has to use some expansions over some of these three parameters. 

The crucial theoretical problems are:
\begin{itemize}
\item 
The development of an effective approach to calculate higher-order 
corrections to the energy levels.
\item
Finding an effective approach to estimate the size of 
uncalculated higher-order corrections to the energy levels. 
\end{itemize}
The difference between these two problems is very important: any particular evaluation can 
include only a part of contributions and we must learn how to determine the uncertainty of the theoretical 
calculation, {\em i. e.} how to estimate corrections that cannot be calculated. 
We discuss below some important higher-order QED corrections, 
our knowledge on which determines the accuracy of 
the bound state QED calculations. Doing {\em ab initio} QED calculations, none can present any 
theoretical prediction to compare with the measurements. With the help of QED one can only express 
some measurable quantities in terms of fundamental constants (like e. g. the Rydberg constant $R_\infty$, 
the fine structure constant $\alpha$, the electron-to-proton ($m_e/m_p$) 
and electron-to-muon ($m_e/m_\mu$) mass ratio). The latter 
have to be determined somehow, however, essential part of the experiments to obtain the values of 
the contstans involves measurements with simple atoms and calculations within bound state QED. 

The study of the Lamb shift in the hydrogen atom and the helium ion 
about fifty years ago led to a great development of Quantum electrodynamics. 
Now, investigations of the lamb shift are still of interest.
After recent calculations of the one-loop, two-loop and three-loop corrections to the Lamb shift 
in the hydrogen atom, the main uncertainty comes from higher order two-loop contributions of the order 
$\alpha^2(Z\alpha)^6m$. They contain logarithms ($\ln(Z\alpha)$) which enhance the  
correction\footnote{In hydrogen, one can find: $\ln^3(Z\alpha)^{-1}\sim 120$, $\ln^2(Z\alpha)^{-1}\sim 24$ 
and $\ln(Z\alpha)^{-1}\sim 5$.}, and the leading term with the cube of the 
logarithm is known \cite{JETP93}. The uncertainty due to the uncalculated next-to-leading terms is estimated 
as 2 ppm.
It is competible with an experimental uncertainty from laser experiments (3 ppm) and essentially 
smaller than the 10-ppm inaccuracy of computations because of the lack of an 
appropriate knowledge of the proton charge 
radius \cite{CJP}. This 10-ppm level of uncertainty due to the proton structure is an obvious 
evidence that the QED is an incomplete theory which deals with photons and leptons (electrons and muons) 
and cannot describe the protons (or deutrons) from first principles. To do any calculations with hydrogen 
and deuterium one has to get some appropriate data on their structure from expreriment.

We particularly discuss here precise tests of the 
bound state QED theory due measurements of the $1s-2s$ and other optical 
transitions in hydrogen, muonium and positronium 
and related experiments. 
We consider a number of different two-photon Doppler-free experiments in hydrogen and deuterium 
in Sect. II. In Sect. III we present the status of the Lamb shift study and discuss some running auxilary 
experiments which can help 
us in the understanding of the higher-order corrections and the proton structure.
Since the problem of the proton structure limits usefulness of extremely precise hydrogen experiments, 
the study of unstable leptonic atoms can provide some tests QED which are
efficient and competitive  with the study of hydrogen \cite{Jungmann}. We discuss these in Sect. IV. 

The paper contains also a brief summary and a list of references. The latter is far incomplete. However,
most problems concerning the precision study of simple atoms were discussed at the recent {\em Hydrogen atom, 2} 
meeting, which took place June, 1-3, 2000, in Italy and we hope that one  
finds more references on the subject therein \cite{h2}.

\section{Two-photon Doppler-free spectroscopy}

The effect of Doppler broadening used to be a limiting factor in the measurement of 
any transition frequency. A 
way to avoid it is to apply two-photon transitions, which are not sensitive to the linear 
Doppler shift. First of all, a success in precision 
spectroscopy is to be expected from $1s-2s$ measurements, because of the metastability 
of the $2s$ state and, hence, of its narrow natural radiative width. Since a value of the 
Rydberg constant can only be determined from measurements with hydrogen and deuterium 
transitions, one has to measure at least two different transitions for any applications to QED 
tests ({\em i. e.} 
one measurement is for the Rydberg constant, while the other is to test QED using 
the value of the Rydberg constant). An appropriate option is to study the $1s-ns$, $2s-ns$ or $2s-nd$ 
transitions in hydrogen, or the $1s-2s$ transition in other atoms.

\subsection{Studying hydrogen and deuterium atoms}

To precisely test bound state QED theory of the Lamb shift one has to measure two different 
transitions in hydrogen and/or deuterium. Combining them properly one can exclude the contribution 
of the Rydberg constant to the transition frequency 
\begin{equation} \label{Schr}
E(nl) = - \frac{c\,h\,R_\infty}{n^2}
\end{equation}
and find a value determined by some known 
relativisitc corrections ($\sim \alpha^2 R_\infty$) and 
by the Lamb shifts of the involved levels. Since a number of different 
states is involved, a number of Lamb shifts have to be determined.   

A specific combination of the Lamb shifts 
\begin{equation} \label{Delt}
\Delta_L (n) = E_L(1s)-n^3E_L(ns)
\end{equation}
is important \cite{JETP94} for the evaluation of the data from optical measurements 
in Garching \cite{Holzwarth} and Paris \cite{Biraben}, obtained by means of two-photon 
Doppler-free laser spectroscopy. The use of this difference allows to present all unknown Lamb 
shifts of $ns$ states in terms of only one of them (usually---either $E_L(1s)$ or $E_L(2s)$).
The uncertainty in this difference is also determined by unknown higher-order two-loop terms, 
but the leading term which includes a squared logarithm is known \cite{JPB}.
It is important 
to underline that the status of this difference \cite{ZP97,CJP} 
differs from the status of the 1s Lamb shift \cite{icap}. 
It is free of most theoretical problems and it is in some sense not a theoretical value, but 
a mathematical one.

\subsubsection*{Absolute measurements of two-photon transitions}

Now let us consider some recent experiments. 
The most accurate result for the Lamb shift 
by optical means (Fig. 1, see \cite{CJP,icap} 
for references)) can be achieved from a comparison of 
the Garching data and the Paris data (see Fig. 2, the references can be found in Refs. 
\cite{CJP,Holzwarth,Biraben,h2}).
These are results after absolutely measuring some transitions frequencies 
({\em i. e.} by measuring with respect to the cesium standard).
In the case of the Garching experiment, a small electric field allows a 
single-photon E1 transition from the 2s state to the 1s level 
and a resonance in the intensity of this decay was used as a signal when tuning the laser frequency. 
The measurement accuracy 
of the 1s-2s transition is high and it can be used for 
other applications, like e. g. a search for variation 
of constants (see e. g. \cite{vari}).

\begin{figure} 
\begin{minipage}[b]{0.45\textwidth}
\epsfig{file=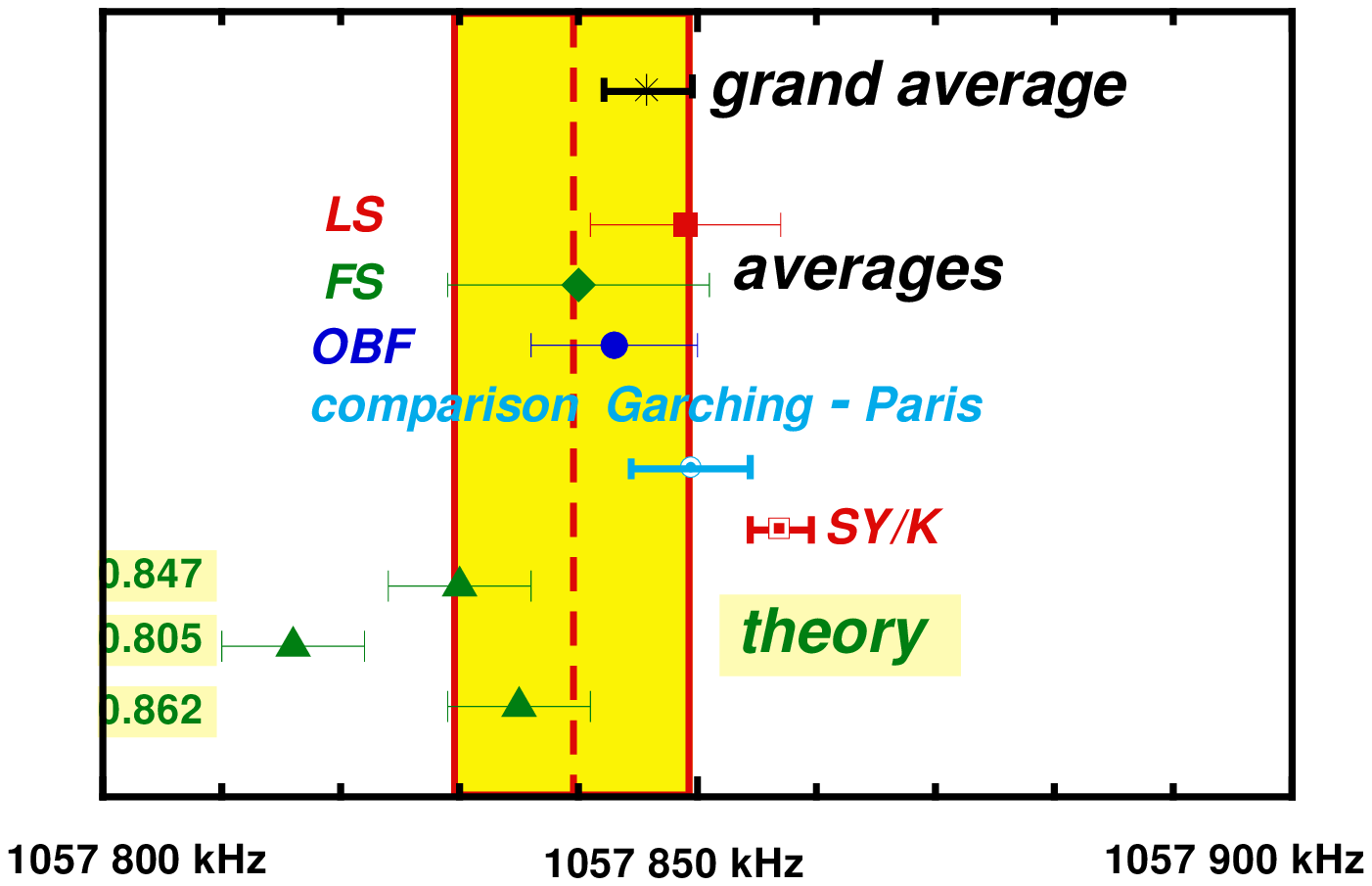,scale=0.52}
\end{minipage}%
\hskip 0.08\textwidth
\begin{minipage}[b]{0.45\textwidth}
\epsfig{file=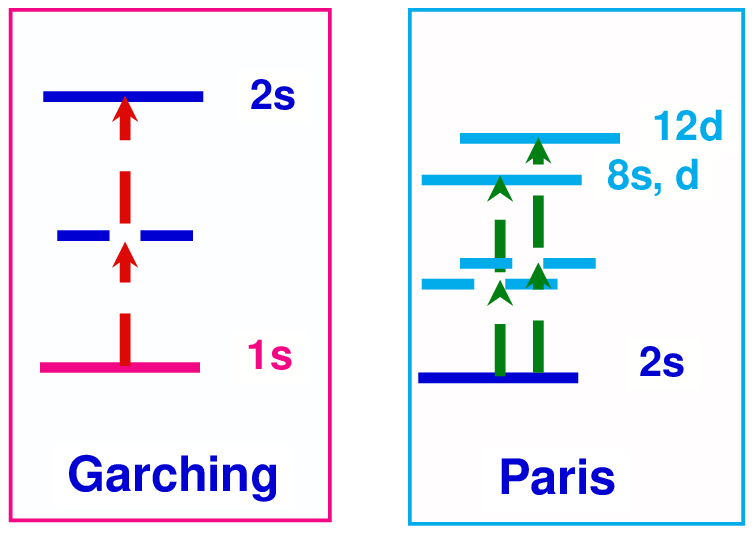,scale=0.9}
\end{minipage}
\vspace{10pt}
\begin{minipage}[t]{0.45\textwidth}
\caption{
$2s-2p$ Lamb shift: comparison of theory and experiment. Grand average denotes the  average of all 
data from the 
Lamb shift (LS), fine structure (FS) and optical beat frequnecy (OBF) measurements.}
\end{minipage}%
\hskip 0.08\textwidth%
\begin{minipage}[t]{0.45\textwidth}
\caption{
Level schemes of absolute frequency measurements at MPQ (Garching) and LKB (Paris)}
\end{minipage}
\end{figure}

\subsubsection*{Relative measurements of two-photon transitions}

A comparison of two absolute frequencies involves cesium standards and a lot of metrology. It is 
possible to avoid comparing two frequencies within the same experiment. Level  
schemes of three experiments (see for detail Refs. \cite{CJP,Holzwarth,Biraben,h2}) are presented in Fig. 3.
Most of them used only two-photon transitions. In all three experiments, the pair of measured frequencies 
consists of two values that differ by a factor either about 4 (Garching, Paris) or 2. 
These factor appear within the leading non-relativistic 
approximation ({\em i. e.} from the Schr\"odinger equation with the Coulomb potential), in which 
the energy levels are determined by Eq. (\ref{Schr}).
Multiplying the 
smaller frequencies by the proper factor (4 or 2) and 
comparing them with the larger frequencies experimentally, the beat frequency signals 
were extracted. For some rather historical reasons, the final results are less accurate 
than in the case of the comparison of two absolute measurements (see Fig. 1 which contains an average 
value over all three experiments). 

\begin{figure}[ht] 
\centerline{\epsfig{file=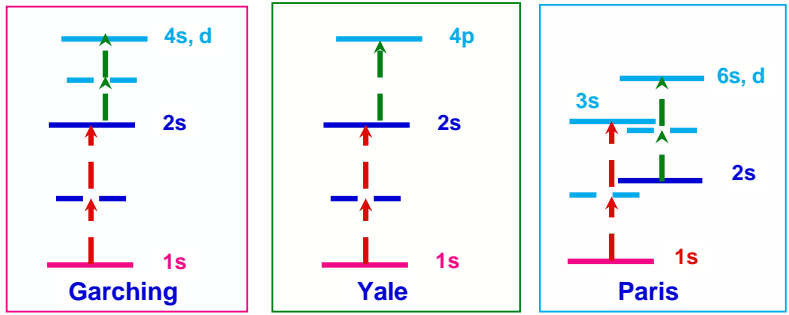,width=0.7\textwidth}}
\vspace{10pt}
\caption{Level schemes of relative frequency measurements at MPQ (Garching), Yale University 
and LKB (Paris)}
\end{figure}

\section{The Lamb shift in the hydrogen atom}

Only a part of the experiments mentioned above were performed for both hydrogen and deuterium, 
and essentially more experimental data are available for hydrogen. 
Below  we discuss only the Lamb shift in the hydrogen 
atom.

\subsection{Present status }

The current situation of the comparison of theory and experiment is summarized 
in Fig. 3 (see Refs. \cite{CJP,icap} 
for details). A result marked with {\em SY/K} (Sokolov-and-Yakovlev 
value, corrected by us) is not included neither in the average over the Lamb shift 
measurements ({\em LS}) nor into the {\em grand 
average}. This corrected result of Sokolov and Yakovlev is claimed to be the most precise, 
however its real accuracy is an open question. The theory is presented with three different values 
of the proton 
size published some time ago. The uncertainty of these theoretical results is about 4 ppm. 
In our opinion, a reasonable theory is not so accurate and its margins are 
presented as a filled area (10-ppm uncertainty). 
All experimental results but one by Sokolov and Yakovlev are consistent with 
our concervative estimate of the theoretical value.

The present status can be briefly described as following:
\begin{itemize}
\item The experiments are mainly consistent to each other and with theory.
\item In particular, the optical data evaluated with the help of the difference in Eq. (\ref{Delt}) 
are consistent 
with the microwave data ({\em LS} and {\em FS} in Fig. 1) found without the use of Eq. (\ref{Delt}).
\item The uncertainty of the grand average value for the $2s-2p_{1/2}$ Lamb shift is about 3 ppm.
\item The computational uncertainty is about 2 ppm and it originates from the unknown higher-order two-loop 
corrections of the order $\alpha^2(Z\alpha)^6m$ which are known only in part \cite{JETP93}.
\item The uncertainty due to the finite size of the proton is about 10 ppm and it is due to 
the inaccuracy 
in the determination of the proton charge radius \cite{CJP}. 
\end{itemize}
Below we discuss some current laser experiments which offer some solutions for the problems with 
the theory, both: of the proton size and of the higher-order two-loop contributions.

\subsection{Proton structure}

There are a few ways to study the proton chagre distribution. One of them is to look for elastic 
scattering of electrons by protons at low momentum transfer $q$. One can determine the proton 
electric form factor from the scattering data and extrapolate in to zero momentum transfer 
\begin{equation}
G_E (q^2) = 1 - \frac{R^2_p}{6} q^2 + ...
\end{equation}
Unfortunately, the scattering data were not evaluated properly and a comprehensive 
 description cannot be available. 
The claimed uncertainty leads to a 3.5-ppm error bar for the Lamb shift, but we expect it to be rather 10-ppm.

A promising project to determine the proton charge radius is now in progress at Paul Scherrer Institut 
(Villigen). It 
deals with muonic hydrogen. The muon is about 200 times heavier than the electron and hence the Bohr orbit 
of the muon lies much lower than the one of the electron and the level energies are
more affected by the proton structure. The used atomic level scheme is presented in Fig. 4. It 
is similar to the one applied for muonic helium some time ago. 
A main advantage is a slow-muon beam at PSI. The use of slow muons allows to make use of 
a low-density gas target which reduces the collisional decay rate of the metastable $2s$ state. 
It has been checked experimentally that, under the conditions of  
the PSI experiment, the $2s$ state is metastable enough and not destroyed by collisions. 
That allows one to go to the 
next step: to excite the atoms in the $2s$ state to the $2p$ state 
by a laser and to look for the intensity of the X-ray Lyman-$\alpha$ as a function 
of the laser frequency. In 
case of successful measurement the result will be the Lamb shift in muonic hydrogen with 
an essential contribution due to the proton size, and eventually with a value of the proton 
charge radius more accurate by an order of magnitude than the current scattering values.
 
\begin{figure}[ht] 
\centerline{\epsfig{file=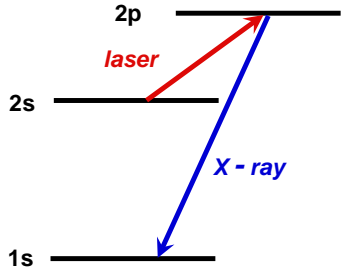,width=0.3\textwidth}}
\vspace{10pt}
\caption{The level scheme used in the 
PSI experiment on the Lamb shift in muonic hydrogen}
\end{figure}

\subsection{Higher-order two-loop corrections}

The other problem of theory of the Lamb shift in hydrogen is the unknown higher-order 
two-loop corrections. They are proportional 
to $Z^6$, while the leading contribution to the Lamb shift is $\sim Z^4$. 
That means that some less accurate 
measurements at higher $Z$ can nevertheless give some 
efficient results for these higher-order terms. 
There are three projects for low-$Z$ ions \cite{h2}:
\begin{itemize}
\item Lamb shift measurement in the $^4$He$^+$ ion ($Z=2$) at the University of Windsor (recently completed);
\item two-photon $2s-3s$ transition in the $^4$He$^+$ ion ($Z=2$) at the University of Sussex (in progress);
\item fine-structure ($2p_{3/2}-2s$) measurement in hydrogen-like nitrogen $^{14}$N$^{6+}$ ($Z=7$) and 
$^{14}$N$^{6+}$ at the Florida State University.
\end{itemize}
The scheme of the last experiment is presented in Fig. 5. It is similar to the previous one with 
the muonic hydrogen. It is expected (see Myers' paper in \cite{h2}) to be sensitive to 
the higher-order two-loop corrections.

\begin{figure}[ht] 
\centerline{\epsfig{file=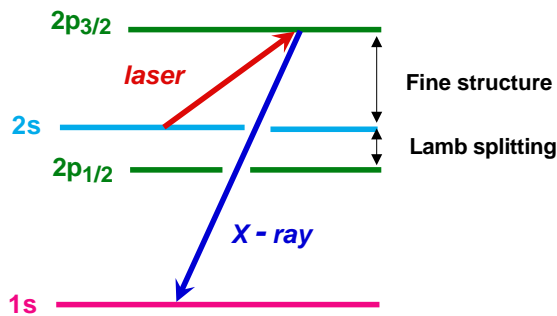,width=0.5\textwidth}}
\vspace{10pt}
\caption{The level scheme of the FSU experiment on the fine structure in hydrogen-like nitrogen}
\end{figure}

\section{Leptonic atoms}

Since the theory of the energy levels in the hydrogen atom is limited by nuclear structure effects,
one can try to study protonless hydrogen-like atoms: muonium and positronium. In both of them 
the $1s-2s$ interval was measured. 

\subsection{1s-2s transition in positronium}

In the positronium spectrum there are a number of values which were or are under 
precise experimental study. In all cases the uncertainty of the positronium energy ($n = 1, 2$) 
is known up to $\alpha^6m$. The only double logarithm ($\alpha^7m\ln^2\alpha$) is known to the next order
\cite{JETP93,Pachucki}. The inaccuracy of the theory originates 
from the non-leading terms (single logarithm and constant) of radiative and radiative-recoil corrections 
of order $\alpha^7m$.
For the majority of measurable quantities (hfs of $1s$, $1s-2s$ interval, 
fine structure of $n=2$, orthopositronium and parapositronium decay rate), 
the theory is competitive with the experiment and actually 
 slightly more accurate. We present a level scheme of a measurement of the 
$1s-2s$ interval in positronium in Fig. 6, 
while in Fig. 7 we compare the theory with the experiment.

\begin{figure} 
\begin{minipage}[b]{0.45\textwidth}
\epsfig{file=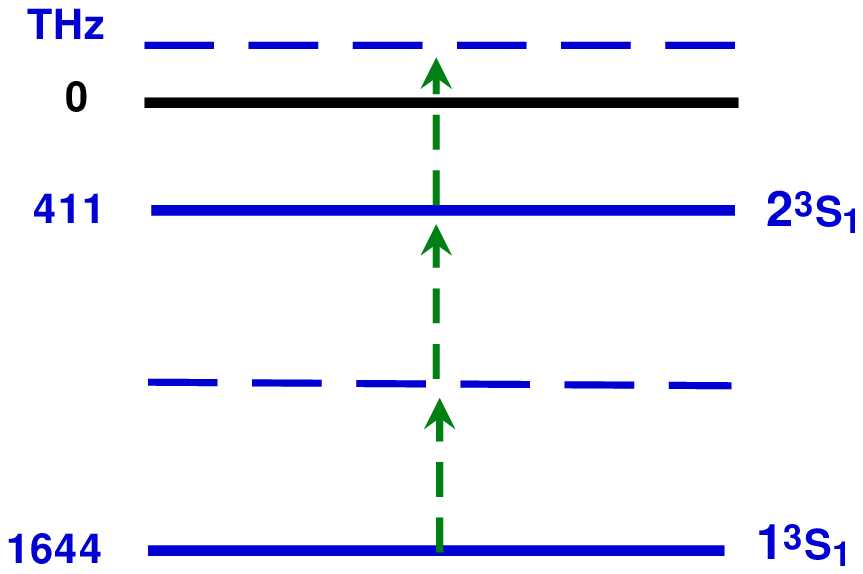,scale=0.7}
\end{minipage}%
\hskip 0.08\textwidth
\begin{minipage}[b]{0.45\textwidth}
\epsfig{file=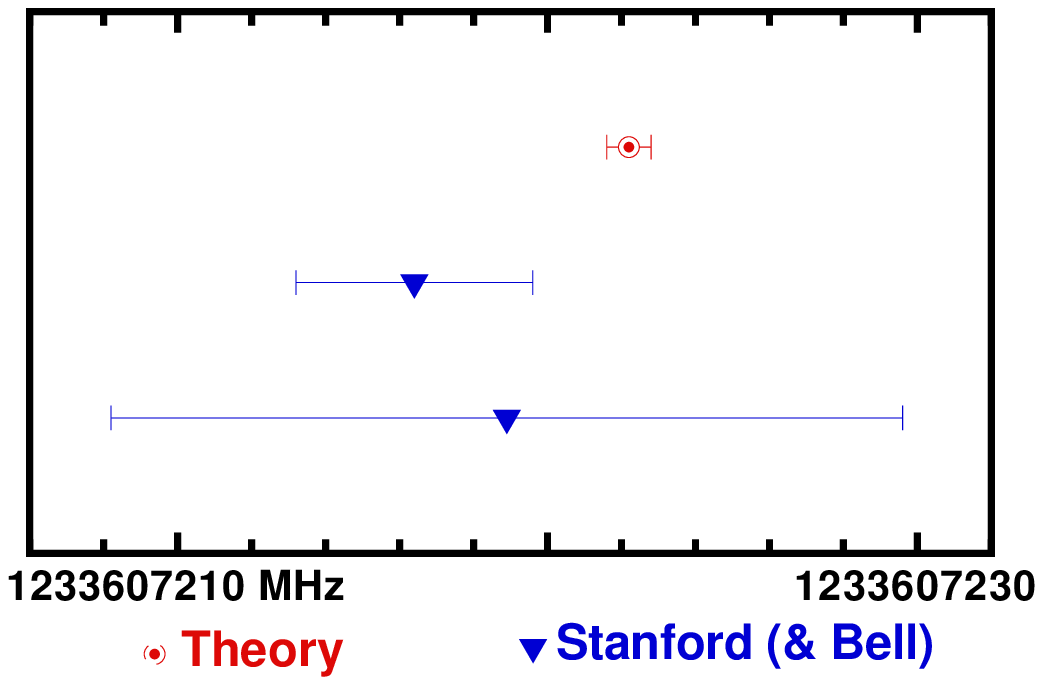,scale=0.6}
\end{minipage}
\vspace{10pt}
\begin{minipage}[t]{0.45\textwidth}
\caption{
Three-photon ionisation in positronium at Stanford University}
\end{minipage}%
\hskip 0.08\textwidth%
\begin{minipage}[t]{0.45\textwidth}
\caption{
Positronium $1s-2s$: theory and experiment}
\end{minipage}
\end{figure}

The positronium $1s-2s$ experiment is quite different from the hydrogen one, because of 
the short lifetime of the 
orthopositronium triplet $1^3S_1$ state which the experiment starts from. The three-photon annihilation 
leads to a lifetime of $1^3S_1$ state of 
$1.4 \cdot 10^{-7}$ s. The method applied was three-photon ionization which has a resonance 
due to the two-photon transition.

\subsection{1s-2s transition in muonium atom}

The muonium nucleus, the muon, lives about 2.2 $\mu$s and 
an idea used to measure $1s-2s$ interval in muonium is similar to the positronium experiment (Fig. 6). 
However, the application is very different. In contrast to 
positronium (in which, e. g. the $1s-2s$ experiment is really competitive with the study of $1s$ hfs), 
a much more efficient test of QED can be provided 
by the hyperfine structure of the ground state which was measured very precisely. 
The uncertainty of the calculation of the $1s$ hfs interval originates from some 
unknown corrections of the fourth order. 
Some of these, 
including the large logarithms ($\ln(Z\alpha)^{-1}\sim 5$ or $\ln(M/m)\sim 5$), 
are known in the double logarithmic approximation \cite{JETP93,ZP96} and non-leading 
terms lead to an uncertainty of 
the theoretical expression as large as 0.05 ppm. The uncertainty arises from the unknown next-to-leading 
radiative-recoil ($\alpha(Z\alpha)^2(m/M)E_F$) and pure recoil ($(Z\alpha)^3(m/M)E_F$)\footnote{
$E_F$ stands here for the Fermi energy, which is a leading order contribution to the hyperfine structure 
which is a result of the non-relativistic interaction of the magnetic moments of the electron and the muon.} 
terms (which are essentially the same as the $\alpha^7m$ terms in positronium). 
However, the budget for the theoretical uncertainty contains not only the computational items. 
Actually, the largest contribution to the budget comes from a calculation of 
the Fermi energy because of the lack of a precise knowledge of the muon-to-electron mass ratio. 
We summarize in Fig. 8 a few of the most accurate values for the mass ratio (see Refs. 
\cite{ZP96,Jungmann} 
for references). The most accurate result there 
is from study of the Zeeman effect of the $1s$ state in muonium. 
Another way to determine this ratio is the $1s-2s$ muonium experiment. 
Two other values are extracted from the study of muon spin precession in 
 different media.

\begin{figure}[ht] 
\centerline{\epsfig{file=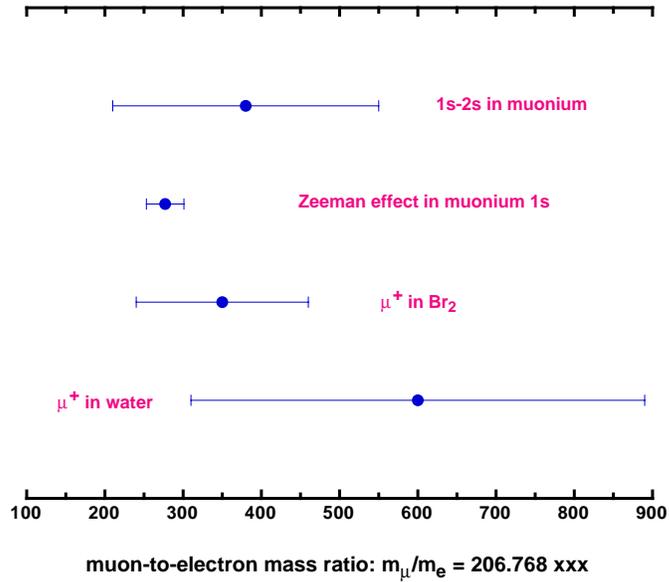,width=0.6\textwidth}}
\vspace{10pt}
\caption{Some determinations of the muon-to-electron mass ratio}
\end{figure}

\section{Summary}

In our talk we briefly discuss several precision tests of bound state QED. The short overview shows 
that theory and experiment are consistent within their uncertainty and the crucial 
corrections in bound state QED in present are:
\begin{itemize}
\item higher order two-loop corrections (hydrogen Lamb shift);
\item radiative-recoil and pure recoil terms of order $\alpha^7m$, the calculation of which 
involves an essential part of the QED, binding and two-body effects (positronium and muonium).
\end{itemize}
The study of these corrections is necessary to develop an efficient theory competitive with experiment.

\section*{Aknowledgements}

I would like to thank T. W. H\"ansch and S. N. Bagayev for support, hospitality and 
stimulating discussions. The stimulating discussions with K. Jungmann, D. Gidley, R. Conti and G. Werth 
are also gratefully acknowledged. I am grateful to J. Reichert for useful remarks. 
The work was supported in part by RFBR (grant \# 00-02-16718), NATO (CRG 960003) and Russian State Program 
{\em Fundamental Metrology}.

\end{document}